\documentclass[11pt,letterpaper]{article}
\usepackage{emnlp2017}
\usepackage{times}
\usepackage{latexsym}

\usepackage{graphicx}
\usepackage{amsmath}
\usepackage{graphicx}
\usepackage{subfigure}
\usepackage{booktabs}

\emnlpfinalcopy

\def\emnlppaperid{***}

\let\svthefootnote\thefootnote
\newcommand\blankfootnote[1]{%
  \let\thefootnote\relax\footnotetext{#1}%
  \let\thefootnote\svthefootnote%
}

\title{Crowdsourcing Multiple Choice Science Questions}

\author{Johannes Welbl\footnotemark[1] \\  Computer Science Department \\ University College London  \\ {\tt j.welbl@cs.ucl.ac.uk}
         \And  Nelson F. Liu\footnotemark[1] \\ Paul G. Allen School of \\ Computer Science \& Engineering \\ University of Washington \\ {\tt nfliu@cs.washington.edu}
         \AND  Matt Gardner \\ Allen Institute for Artificial Intelligence \\ {\tt mattg@allenai.org}
         }
         
\date{}

\begin{document}
\maketitle
\blankfootnote{\llap{\textsuperscript{*}}Work done while at the Allen Institute for Artificial Intelligence.}
\begin{abstract}
  
  We present a novel method for obtaining high-quality, domain-targeted multiple choice questions from crowd workers. 
  Generating these questions can be difficult without trading away originality, relevance or diversity in the  answer options.
  Our method addresses these problems by leveraging a large corpus of domain-specific text and a small set of existing questions. It produces model suggestions for document selection and answer distractor choice which aid the human question generation process.
  With this method we have assembled \emph{SciQ}, a dataset of 13.7K
  multiple choice science exam questions.\footnote{Dataset available at
  \url{http://allenai.org/data.html}
  } 
  We demonstrate that the method produces in-domain questions by providing an analysis of this new dataset and by showing that humans cannot distinguish the crowdsourced questions from original questions. 
  When using \emph{SciQ} as additional training data to existing questions, we observe accuracy improvements on real science exams.

\end{abstract}

\begin{figure*}
	\centering
	\includegraphics[width=1.0\linewidth]{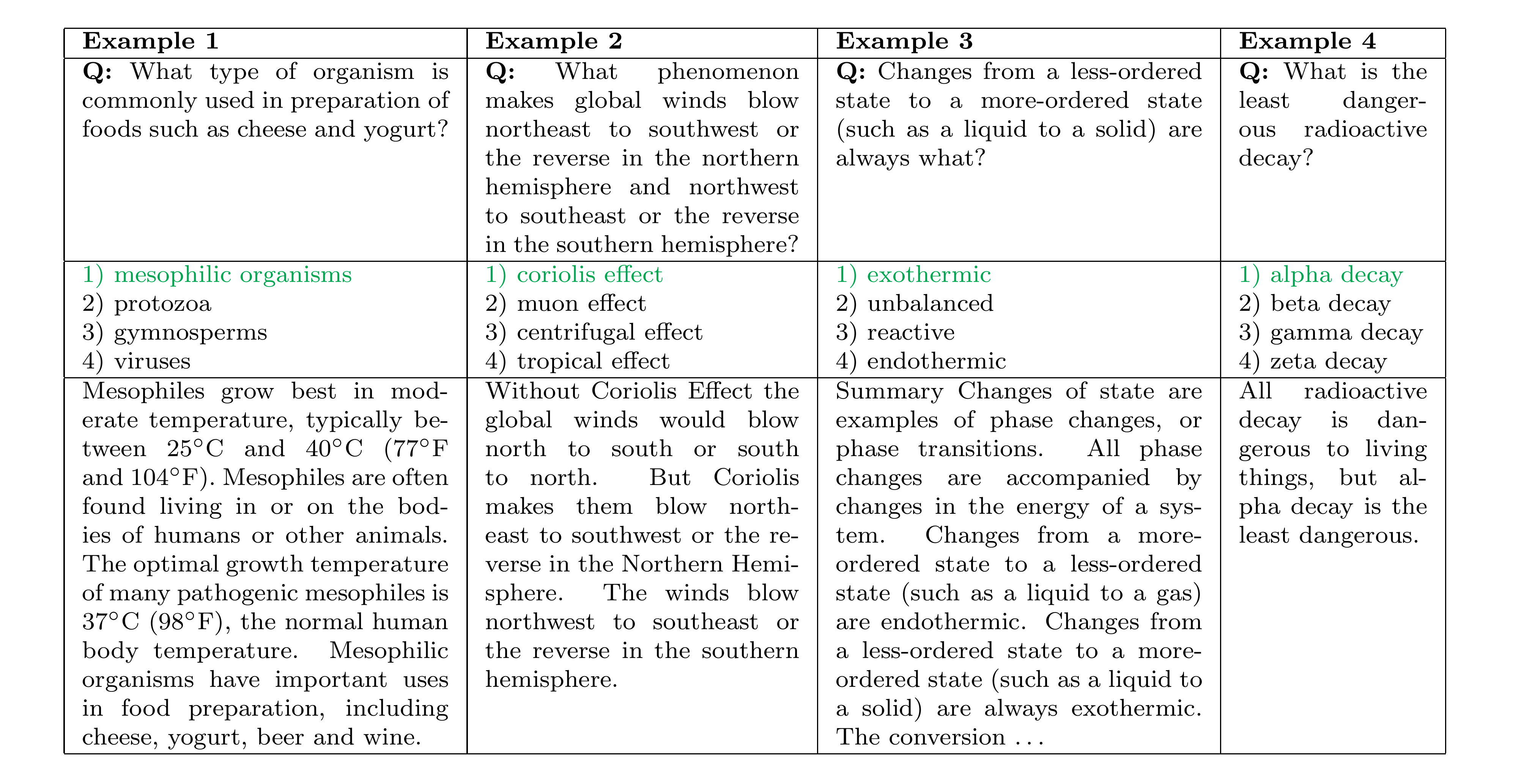}
    \caption{The first four \emph{SciQ} training set examples. An instance consists of a question and 4 answer options (the correct one in green). Most instances come with the document used to formulate the question.}
	\label{fig:examples}
\end{figure*}

\section{Introduction}

The construction of large, high-quality datasets has been one of the main drivers of progress in NLP. 
The recent proliferation of datasets for textual entailment, reading comprehension and Question Answering (QA) ~\cite{Bowman2015SNLI,%
Hermann2015News,rajpurkar2016SQUAD,Hill2015CBT,Hewlett2016WikiReading,Nguyen2016MSMARCO} has allowed for advances on these tasks, particularly with neural
models~\cite{Kadlec2016ASRNetwork,Dgingra2016Gated,Sordoni2016Iterative,Seo2016Bidirectional}.
These recent datasets cover broad and general domains, but progress on these datasets has not translated into similar improvements in more targeted domains, such as science exam QA.

Science exam QA is a high-level NLP task which requires the mastery and integration of information extraction, reading comprehension and common sense reasoning \cite{Clark2013Study,Clark2015Elementary}. 
Consider, for example, the question \emph{``With which force does the moon affect tidal movements of the oceans?''}.
To solve it, a model must possess an abstract understanding of natural phenomena and apply it to new questions. 
This transfer of general and domain-specific background knowledge into new scenarios poses a formidable challenge, one which modern statistical techniques currently struggle with.  In a recent Kaggle competition addressing 8\textsuperscript{th} grade science questions~\cite{schoenick2016moving}, the highest scoring systems achieved only 60\% on a multiple choice test, with retrieval-based systems far outperforming neural systems.

A major bottleneck for applying sophisticated statistical techniques to science QA is the lack of large in-domain training sets.  Creating a large, multiple choice science QA dataset is challenging, since crowd workers cannot be expected to have domain expertise, and questions can lack relevance and diversity in structure and content. Furthermore, poorly chosen answer distractors in a multiple choice setting can make questions almost trivial to solve.

The first contribution of this paper is a general method for mitigating the difficulties of crowdsourcing QA data, with a particular focus on multiple choice science questions. The method is broadly similar to other recent work~\cite{rajpurkar2016SQUAD},
relying mainly on showing crowd workers a passage of text and having them ask a question about it.
However, unlike previous dataset construction tasks, we (1) need domain-relevant passages and questions,
and (2) seek to create multiple choice questions, not direct-answer questions. 

We use a two-step process to solve these problems, 
first using a noisy classifier to find relevant passages and
showing several options to workers to select from when generating a question.  Second, we use a
model trained on real science exam questions to predict good answer distractors given a question and a
correct answer. We use these predictions to aid crowd workers in transforming the question produced from the
first step into a multiple choice question.
Thus, with our methodology we leverage existing study texts and science
questions to obtain new, relevant questions and plausible answer distractors. Consequently, the human
intelligence task is shifted away from a purely \emph{generative} task (which is slow,
difficult, expensive and can lack diversity in the outcomes when repeated) and reframed in terms of
a \emph{selection}, \emph{modification} and \emph{validation} task (being faster, easier, cheaper
and with content variability induced by the suggestions provided). 

The second contribution of this paper is a dataset constructed by following this methodology.  With
a total budget of \$10,415, we collected 13,679 multiple choice science questions, which we call
\emph{SciQ}. Figure \ref{fig:examples} shows the first four training examples in \emph{SciQ}.  This dataset
has a multiple choice version, where the task is to select the
correct answer using whatever background information a system can find given a question and several answer options, and a direct answer
version, where given a passage and a question a system must predict the span within the passage
that answers the question.  With experiments using recent state-of-the-art reading comprehension
methods, we show that this is a useful dataset for further research.  Interestingly, neural models
do not beat simple information retrieval baselines on the multiple choice version of this dataset,
leaving room for research on applying neural models in settings where training examples number in
the tens of thousands, instead of hundreds of thousands.  We also show that using \emph{SciQ} as an
additional source of training data improves performance on real 4\textsuperscript{th} and
8\textsuperscript{th} grade 
exam questions, proving that our method successfully produces useful in-domain training data.

\section{Related Work}

\textbf{Dataset Construction.} A lot of recent work has focused on constructing large datasets
suitable for training neural models.  QA datasets have been assembled based on
Freebase~\cite{Berant13WebQuestions,Bordes2015SimpleQuestions}, Wikipedia
articles~\cite{Yi2015WikiQA,rajpurkar2016SQUAD,Hewlett2016WikiReading} and web search user queries
\cite{Nguyen2016MSMARCO}; for reading comprehension (RC) based on news~\cite{Hermann2015News,Onishi2016Who}, children books~\cite{Hill2015CBT} and
novels~\cite{paperno2016lambada}, and for recognizing textual entailment
based on image captions~\cite{Bowman2015SNLI}.
We continue this line of work and construct a dataset for science exam QA. Our
dataset differs from some of the aforementioned datasets in that it consists of natural language
questions produced by people, instead of cloze-style questions.  It also differs
from prior work 
in that we aim at the narrower
domain of science exams and in that we produce multiple choice questions, which are more difficult to generate.

\textbf{Science Exam Question Answering.} Existing models for multiple-choice science
exam QA vary in their reasoning framework and training methodology. A set of sub-problems and solution strategies are outlined in \newcite{Clark2013Study}.
The method described by \newcite{Li2015Answering} evaluates the coherence of a scene constructed
from the question enriched with background KB information, while~\newcite{Sachan2016Instructional}
train an entailment model that derives the correct answer from background
knowledge aligned with a max-margin ranker. Probabilistic reasoning approaches include Markov
logic networks~\cite{Khot2015MLN} and an integer linear program-based model that assembles proof chains over structured knowledge~\cite{Khot2016TableILP}.  The Aristo ensemble 
\cite{Clark2016Combining} combines multiple reasoning strategies with shallow statistical
methods based on lexical co-occurrence and IR, which by themselves provide surprisingly strong
baselines.  There has not been much work applying neural networks to this task, likely because
of the paucity of training data; this paper is an attempt to address this issue by constructing a
much larger dataset than was previously available, and we present results of experiments using
state-of-the-art reading comprehension techniques on our datasets.

\textbf{Automatic Question Generation.} Transforming text into questions has been tackled before, mostly for didactic purposes. 
Some approaches rely on syntactic transformation
templates~\cite{Mitkov2003ComputerAided,Heilman2010GoodQuestion}, while most others generate cloze-style questions.  
Our first attempts at constructing a science question dataset followed these techniques. We found the methods did not produce high-quality science questions, as
there were problems with selecting relevant text, generating reasonable distractors,
and formulating coherent questions.

Several similarity measures have been employed for selecting answer
distractors~\cite{Mitkov2009Semantic}, including measures derived from
WordNet~\cite{Mitkov2003ComputerAided}, thesauri~\cite{Sumita2005Measuring} and distributional
context~\cite{Pino2008Selection,Aldabe2010Automatic}. Domain-specific ontologies~\cite{Papasalouros2009Automatic},
phonetic or morphological similarity~\cite{Pino2009Semi-automatic,Correia2010Automatic},
probability scores for the question context~\cite{Mostow2012Diagnostic} and context-sensitive
lexical inference~\cite{Zesch2014Automatic} have also been used.
In contrast to the aforementioned similarity-based selection strategies, our method uses a
feature-based ranker to learn plausible distractors from original questions. Several of
the above heuristics are used as features in this
ranking model.  Feature-based distractor generation models~\cite{Sakaguchi2013Discriminative} have
been used in the past by~\newcite{Agarwal2011Automatic} for creating biology questions. Our model
uses a random forest to rank candidates; it is agnostic towards taking cloze or humanly-generated
questions, and it is learned specifically to generate distractors that resemble those in
real science exam questions.

\section{Creating a science exam QA dataset}
\label{sec:method}

In this section we present our method for crowdsourcing science exam questions. The method is a
two-step process: first we present a set of candidate passages to a crowd worker, letting the
worker choose one of the passages and ask a question about it.  Second, another worker takes the
question and answer generated in the first step and produces three distractors, aided by a model
trained to predict good answer distractors.  The end result is a multiple choice science question,
consisting of a question $q$, a passage $p$, a correct answer $a^*$, and a set of distractors, or
incorrect answer options, $\{a'\}$.  Some example questions are shown in
Figure~\ref{fig:examples}.
The remainder of this section
elaborates on the two steps in our question generation process.

\subsection{First task: producing in-domain questions}

Conceiving an original question from scratch in a specialized domain is surprisingly difficult;
performing the task repeatedly involves the danger of falling into specific lexical and structural
patterns.  To enforce diversity in question content and lexical expression, and to inspire relevant in-domain questions,
we rely on a corpus of in-domain text about which crowd workers ask questions.  However, not all text in a large in-domain corpus, such as a textbook, is suitable for generating questions.
We use a simple filter to narrow down the selection to paragraphs likely to produce reasonable questions.

\textbf{Base Corpus.} Choosing a relevant, in-domain base corpus to inspire the questions is of
crucial importance for the overall characteristics of the dataset. For science questions, the
corpus should consist of topics covered in school exams, but not be too linguistically complex,
specific, or loaded with technical detail (e.g., scientific papers).  We observed that articles
retrieved from web searches for science exam keywords (e.g. ``animal'' and ``food'') yield a
significant proportion of commercial or otherwise irrelevant documents and did not consider this
further.  Articles from science-related categories in Simple Wikipedia are more targeted and factual,
but often state highly specific knowledge (e.g., ``Hoatzin can reach 25 inches in length and 1.78
pounds of weight.'').

We chose science study textbooks as our base corpus because they are directly relevant and
linguistically tailored towards a student audience. They contain verbal descriptions of general
natural principles instead of highly specific example features of particular species. While the
number of resources is limited, we compiled a list of 28 books from various online learning
resources, including CK-12\footnote{\url{www.ck12.org}} and
OpenStax\footnote{\url{www.openstax.org}}, who share this material under a Creative Commons
License. The books are about biology, chemistry, earth science and physics and span elementary
level to college introductory material. A full list of the books we used can be found in the
appendix. 

\textbf{Document Filter.} 
We designed a rule-based document filter model into which individual paragraphs of the base corpus are fed. The system classifies individual sentences and accepts a paragraph if a minimum number of sentences is accepted. With a small manually annotated dataset of sentences labelled as either relevant or irrelevant, the filter was designed iteratively by adding filter rules to first improve precision and then recall on a held-out validation set. 
The final filter included lexical, grammatical, pragmatical and complexity based rules. Specifically, sentences were filtered out if they \emph{i)} were a question or exclamation \emph{ii)} had no verb phrase \emph{iii)} contained modal verbs \emph{iv)} contained imperative phrases \emph{v)} contained demonstrative pronouns \emph{vi)} contained personal pronouns other than third-person \emph{vii)} began with a pronoun \emph{viii)} contained first names \emph{ix)} had less than 6 or more than 18 tokens or more than 2 commas \emph{x)} contained special characters other than punctuation \emph{xi)} had more than three tokens beginning uppercase \emph{xii)} mentioned a graph, table or web link \emph{xiii)} began with a discourse marker (e.g.~\emph{`Nonetheless'})  \emph{xiv)} contained absoulute wording (e.g.~\emph{`never', `nothing', `definitely'}) \emph{xv)} contained instructional vocabulary ( \emph{`teacher'}, \emph{`worksheet'}, \dots).

Besides the last, these rules are all generally applicable in other domains to identify simple declarative statements in a corpus.

\textbf{Question Formulation Task.} 
To actually generate in-domain QA pairs, we
presented the filtered, in-domain text to crowd workers and had them ask a question that could be answered by the presented passage. Although most undesirable paragraphs had been filtered out beforehand, a non-negligible proportion of irrelevant documents remained. To circumvent this problem, we showed each worker \emph{three} textbook paragraphs and gave them the freedom
to choose one or to reject all of them if irrelevant. Once a paragraph had been chosen, it was not reused to formulate more questions about it. We further specified desirable characteristics of science exam questions: 
no \emph{yes/no} questions, not requiring further context, querying general principles rather than highly specific facts, question length between 6-30 words, answer length up to 3 words (preferring shorter), no ambiguous questions, answers clear from paragraph chosen. Examples for both desirable and undesirable questions were
given, with explanations for why they were good or bad examples. 
Furthermore we encouraged workers to give feedback, and a contact email was provided to address upcoming questions directly; multiple crowdworkers made use of this opportunity. The task was advertised on \emph{Amazon Mechanical Turk}, requiring \emph{Master's} status for the crowdworkers,
and paying a compensation of 0.30\$ per HIT. A total of 175 workers participated in the whole crowdsourcing project.

In 12.1\% of the cases all three documents were rejected, much fewer than if a single document had been presented (assuming the same proportion of relevant documents).
Thus, besides being more economical,
proposing several documents reduces the risk of generating irrelevant questions and in the best case helps match a crowdworker's individual preferences.

\subsection{Second task: selecting distractors}

Generating convincing answer distractors is of great importance, since bad distractors can make a question trivial to solve. When writing science
questions ourselves, we found that finding reasonable distractors was the most time-consuming
part overall.  Thus, we support the process in our crowdsourcing task with model-generated answer distractor suggestions. This primed the workers with 
relevant examples, and we allowed them to use the suggested distractors directly if they were good enough. We next discuss characteristics of good answer distractors, 
propose and evaluate a model for suggesting such distractors, and describe the crowdsourcing task that uses them.

\textbf{Distractor Characteristics.} Multiple choice science questions with nonsensical incorrect
answer options are not interesting as a task to study, nor are they useful for training a model to
do well on real science exams, as the model would not need to do any kind of science reasoning to
answer the training questions correctly.  The difficulty in generating a good multiple choice question,
then, lies not in identifying expressions which are false answers to $q$, but in generating
expressions which are \emph{plausible} false answers. Concretely, besides being false answers, good distractors should thus:
\begin{itemize}
    \itemsep0em
    \item be grammatically consistent: for the question ``When animals use energy, what is always
      produced?'' a noun phrase is expected.
    \item be consistent with respect to abstract properties: if the correct answer belongs to a
      certain category (e.g., chemical elements) good distractors likely should as well.
    \item be consistent with the semantic context of the question: a question about animals and energy
    should not have \emph{newspaper} or \emph{bingo} as distractors.
\end{itemize}

\textbf{Distractor Model Overview.} We now introduce a model which generates plausible answer distrators and takes into account the above criteria. On a basic level, it ranks candidates from a large collection $C$ of possible distractors and selects the highest scoring items. Its ranking function
\begin{equation}
r: (q,a^*,a') \mapsto s_{a'} \in [0,1]
\end{equation}
produces a confidence score $s_{a'}$ for whether $a' \in C$ is a good distractor in the
context of question $q$ and correct answer $a^*$. 
For $r$ we use the scoring function $s_{a'} = P(a'\ \text{is good } |~q,a^*)$ of a binary classifier which distinguishes plausible (good) distractors from random (bad) distractors based on features $\phi(q,a^*,a')$. 
For classification, we train $r$ on actual in-domain questions with observed false answers as the plausible (good) distractors, and random expressions as negative examples, sampled in equal proportion from $C$. As classifier we chose a random forest~\cite{Breiman2001Random}, because of its robust performance in small and mid-sized data settings and its power to incorporate nonlinear feature interactions, in contrast, e.g., to logistic regression.

\textbf{Distractor Model Features.} This section describes the features $\phi(q,a^*,a')$ used by the distractor ranking model. With these features, the distractor model can learn characteristics of real distractors from original questions and will suggest those distractors that it deems the most realistic for a question. 
The following features of question $q$, correct answer $a^*$ and a tentative distractor expression $a'$ were used:

\begin{itemize}
\item  bags of $GloVe$ embeddings for $q$, $a^*$ and $a'$; 
\item  an indicator for POS-tag consistency of $a^*$ and $a'$;
\item  singular/plural consistency of $a^*$ and $a'$;
\item  log.~avg.~word frequency in $a^*$ and $a'$;
\item  Levenshtein string edit distance between $a^*$ and $a'$; 
\item  suffix consistency of $a^*$ and $a'$ (firing e.g.~for (\emph{regeneration, exhaustion})); 
\item  token overlap indicators for $q$, $a^*$ and $a'$;    
\item  token and character length for $a^*$ and $a'$ and similarity therein;  
\item  indicators for numerical content in $q$, $a^*$ and $a'$ consistency therein;
\item  indicators for units of measure in $q$, $a^*$ and $a'$, and for co-occurrence of the same unit; 
\item  WordNet-based hypernymy indicators between tokens in $q$, $a^*$ and $a'$, in both directions and potentially via two steps; 
\item  indicators for 2-step connections between entities in $a^*$ and $a'$ via a KB based on OpenIE triples \cite{Mausam2012OpenIE} extracted from pages in Simple Wikipedia about anatomical structures;
\item  indicators for shared Wordnet-hyponymy of $a^*$ and $a'$ to one of the concepts most frequently generalising all three question distractors in the training set (e.g.~\emph{element}, \emph{organ}, \emph{organism}). 
\end{itemize}

The intuition for the knowledge-base link and hypernymy indicator features is that they can reveal sibling structures of $a^*$ and $a'$ with respect to a shared property or hypernym. For example, if the correct answer $a^*$ is \emph{heart}, then a plausible distractor $a'$ like \emph{liver} would share with $a^*$ the hyponymy relation to \emph{organ} in WordNet.

\textbf{Model Training.} 
We first constructed a large candidate distractor set $C$ whose items were to be ranked by the model. 
$C$ contained 488,819 expressions, consisting of
(1) the 400K items in the GloVe vocabulary~\cite{pennington2014glove}; (2) answer
distractors observed in training questions; (3) a list of noun phrases
from Simple Wikipedia articles about body parts; (4) a noun vocabulary of $\mathtt{\sim}$6000 expressions extracted from primary school science texts. 
In examples where $a^*$ consisted of
multiple tokens, we added to $C$ any expression that could be obtained by exchanging one unigram in $a^*$ with another unigram from $C$. 

The model was then trained on a set of 3705 science exam
questions ($4^{\text{th}}$ and $8^{\text{th}}$ grade)
, separated into $80\%$ training questions and $20\%$ validation questions. 
Each question came with four answer options, providing three good distractor examples.  We used \texttt{\small{scikit-learn}}'s
implementation of random forests with default parameters.  We used 500 trees and enforced at least 4 samples per tree leaf.

\textbf{Distractor Model Evaluation.}  Our model achieved $99,4\%$ training and $94,2\%$ validation accuracy overall. 
Example predictions of the distractor model are shown in Table \ref{tab:candidates}. Qualitatively, the
predictions appear acceptable in most cases, though the quality is not high
enough to use them directly without additional filtering by crowd workers.  In many cases the distractor is
semantically related, but does not have the correct type
(e.g., in column 1, ``nutrient'' and ``soil'' are not
elements). Some predictions are misaligned in their level of specificity (e.g.~``frogs'' in column 3), and multiword expressions were more likely to be unrelated or ungrammatical despite the inclusion of part of speech features.

Even where the predicted distractors are not fully coherent, showing them to a crowd worker still has a positive priming effect, helping the worker
generate good distractors either by providing nearly-good-enough candidates, or by forcing
the worker to think why a suggestion is not a good distractor for the question.

\begin{table*}[ht]
\footnotesize
\begin{center}
\begin{tabular}{|p{.25\textwidth}|p{.2\textwidth}|p{.2\textwidth}|p{.25\textwidth}|}
\hline
{\bfseries Q:} Compounds containing an atom of what element, bonded in a hydrocarbon framework, are classified as amines?  & 
{\bfseries Q:} Elements have orbitals that are filled with what?   &
{\bfseries Q:} Many species use their body shape and coloration to avoid being detected by what? &
{\bfseries Q:} The small amount of energy input necessary for all chemical reactions to occur is called what? 
\\ \hline 
{\bfseries A:} nitrogen     & {\bfseries A:} electrons      & {\bfseries A:} predators      & {\bfseries A:} activation energy\\
\hline
{\bfseries oxygen} (0.982)      & {\bfseries ions} (0.975)  & {\bfseries viruses} (0.912)   & conversely energy (0.987) \\
{\bfseries hydrogen} (0.962)    & atoms (0.959)     & ecosystems (0.896)    & {\bfseries decomposition energy} (0.984)  \\
nutrient (0.942)    & crystals (0.952)  & frogs (0.896)         & membrane energy (0.982)       \\
calcium (0.938)     & protons (0.951)   & distances (0.8952)    & motion energy (0.982)         \\
silicon (0.938)     & neutrons (0.946)  & {\bfseries males} (0.877)         & context energy (0.981)        \\
soil (0.9365)       & {\bfseries photons} (0.912)   & crocodiles (0.869)    & {\bfseries distinct energy} (0.980)       \\
\hline

\end{tabular}
\end{center}
\caption{\label{tab:candidates} Selected distractor prediction model outputs. For each QA pair, the top six predictions are listed in row 3 (ranking score in parentheses). Boldfaced candidates were accepted by crowd workers.}
\end{table*}

\textbf{Distractor Selection Task.} To actually generate a multiple choice science question, we
show the result of the first task, a $(q, a^*)$ pair, to a crowd worker, along with the top six distractors suggested from the previously described model. The goal of this task is two-fold: (1) quality control (validating
a previously generated $(q, a^*)$ pair), and (2) validating the predicted distractors or writing
new ones if necessary.

The first instruction was to judge whether the question could appear in a school science exam;
questions could be marked as ungrammatical, having a false answer, being unrelated to science or
requiring very specific background knowledge.  The total proportion of questions passing was
$92.8\%$.

The second instruction was to select up to two of the six suggested distractors, and to write at
least one distractor by themselves such that there is a total of three. The requirement for the
worker to generate one of their own distractors, instead of being allowed to select three predicted
distractors, was added after an initial pilot of the task, as we found that it forced workers to 
engage more with the task and resulted in higher quality distractors.

We gave examples of desirable and undesirable distractors and the opportunity to provide feedback,
as before. We advertised the task on \emph{Amazon Mechanical Turk}, paying 0.2\$ per HIT, again
requiring AMT \emph{Master's} status.  On average, crowd workers found the predicted distractors
good enough to include in the final question around half of the time, resulting in 36.1\% of the
distractors in the final dataset being generated by the model (because workers were only allowed to
pick two predicted distractors, the theoretical maximum is 66\%).  Acceptance rates were
higher in the case of short answers, with almost none accepted for the few cases with very long
answers.

The remainder of this paper will investigate properties of \emph{SciQ}, the dataset we generated by following the methodology described in this section.
We present system and human performance, and we show that \emph{SciQ} can be used as additional training data to improve model performance on real science exams.

\begin{figure}[ht]
  \centering
  \includegraphics[width=0.8\linewidth]{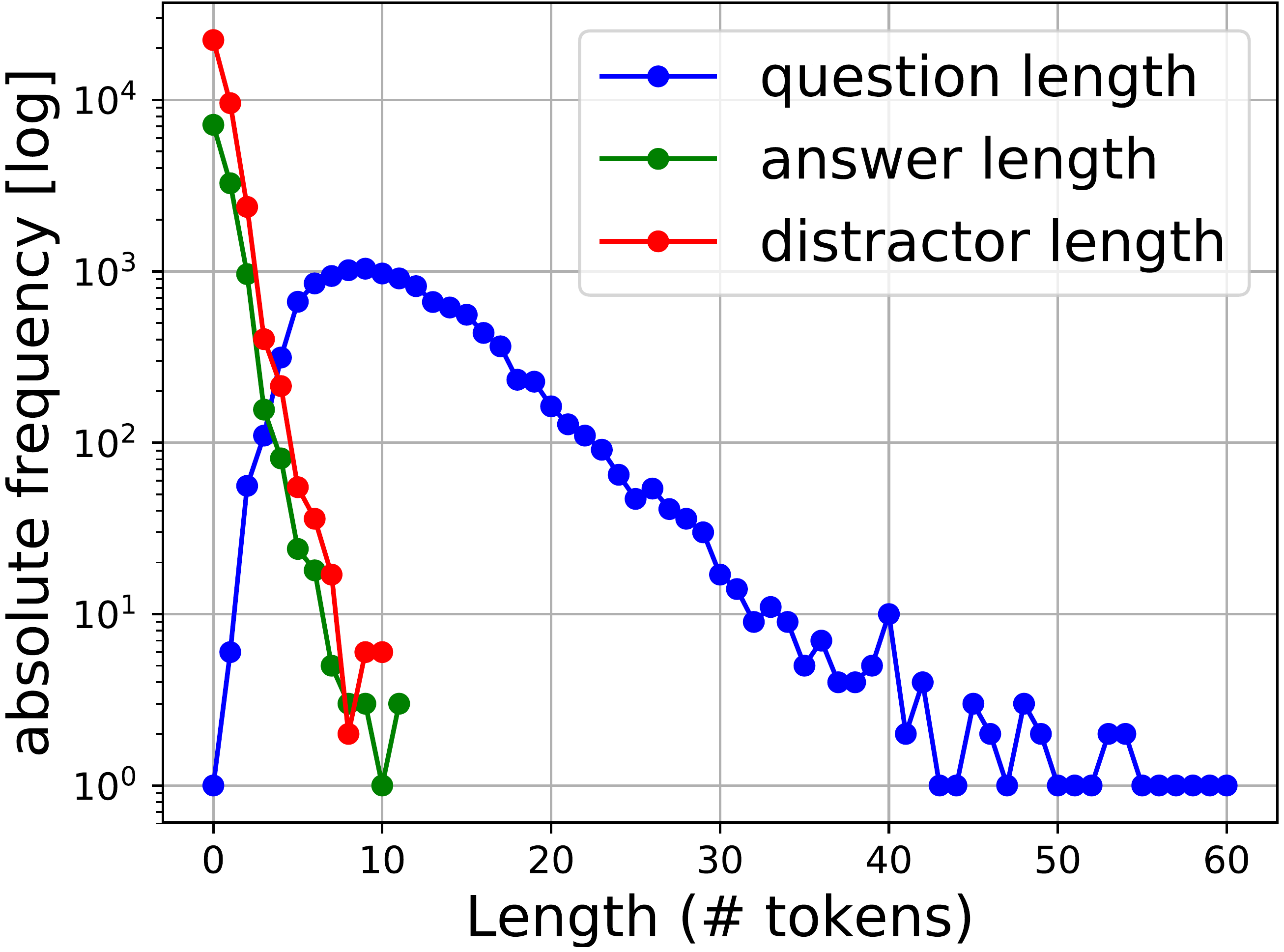}
  \caption{Total counts of question, answer and distractor length, measured
  in number of tokens, calculated across the training set.}
  \label{fig:lengths}
\end{figure}

\begin{table}[ht]
  \centering
  \begin{tabular}{lr}
    \toprule
    \textbf{Model} & \textbf{Accuracy} \\
    \midrule
    Aristo & 77.4 \\
    Lucene & 80.0 \\
    TableILP & 31.8 \\
    \midrule
    AS Reader & 74.1 \\
    GA Reader & 73.8 \\
    \midrule
    Humans & 87.8 $\pm$ 0.045 \\
    \bottomrule
  \end{tabular}
  \caption{Test set accuracy of existing models on the multiple choice version of \emph{SciQ}.}
  \label{tab:mc-performance}
\end{table}

\subsection{Dataset properties}

\emph{SciQ} has a total of 13,679 multiple choice questions.  We randomly shuffled this dataset and
split it into training, validation and test portions, with 1000 questions in each of the validation
and test portions, and the remainder in train.  In Figure \ref{fig:lengths} we show the
distribution of question and answer lengths in the data. For the most part, questions and answers
in the dataset are relatively short, though there are some longer questions.

Each question also has an associated passage used when generating the question.  Because the
multiple choice question is trivial to answer when given the correct passage, the multiple choice
version of \emph{SciQ} does not include the passage; systems must retrieve their own background
knowledge when answering the question.  Because we have the associated passage, we
additionally created a direct-answer version of \emph{SciQ}, which has the passage and the question,
but no answer options.  A small percentage of the passages were obtained from unreleasable texts, so the direct answer version of \emph{SciQ} is slightly smaller, with 10481
questions in train, 887 in dev, and 884 in test.

\textbf{Qualitative Evaluation.} We created a crowdsourcing task with the following
setup: A person was presented with an original science exam question and a crowdsourced question.
The instructions were to choose which of the two questions was more likely to be the real exam question. We randomly drew 100 original questions and 100 instances from the \emph{SciQ} training
set and presented the two options in random order. People identified the science exam question in
55\% of the cases, which falls below the significance level
of p=0.05 under a null hypothesis of a random guess\footnote{using normal approximation}.

\section{SciQ Experiments}

\subsection{System performance}

We evaluated several state-of-the-art science QA systems, reading comprehension
models, and human performance on \emph{SciQ}.

\textbf{Multiple Choice Setting.}  We used the Aristo
ensemble~\cite{Clark2016Combining}, and two of its individual components: 
a simple information retrieval baseline (Lucene), and a table-based integer linear
programming model (TableILP), to evaluate \emph{SciQ}.  
We also evaluate two competitive neural reading comprehension models:
the Attention Sum Reader (AS Reader, a GRU with a pointer-attention
mechanism; \newcite{Kadlec2016ASRNetwork}) and the Gated Attention Reader (GA Reader, an AS Reader with
additional gated attention layers; \newcite{Dgingra2016Gated}).  These reading comprehension methods
require a supporting text passage to answer a question.  We use the same corpus as Aristo's
Lucene component to retrieve a text passage, by formulating five queries based on the question
and answer\footnote{The question text itself, plus each of the four answer options appended to the
question text.} and then concatenating the top three results from each query into a passage.  We train
the reading comprehension models on the training set with hyperparameters recommended by prior
work (\cite{Onishi2016Who} for the AS Reader and \cite{Dgingra2016Gated} for the GA Reader), with
early stopping on the validation data\footnote{For training and hyperparameter details,
see Appendix}.  Human accuracy is estimated using a sampled subset of 650 questions, with 13
different people each answering 50 questions.  When answering the questions,
people were allowed to query the web, just as the systems were.

Table~\ref{tab:mc-performance} shows the results of this evaluation.  Aristo performance is
slightly better on this set than on real science exams (where Aristo achieves 71.3\%
accuracy~\cite{Clark2016Combining}).\footnote{We did not retrain the Aristo ensemble for
\emph{SciQ}; it might overly rely on TableILP, which does not perform well here.}  Because
TableILP uses a hand-collected set of background knowledge that does not cover the topics in
\emph{SciQ}, its performance is substantially worse here than on its original test set.  Neural
models perform reasonably well on this dataset, though, interestingly, they are not able to
outperform a very simple information retrieval baseline, even when using exactly the same
background information.  This suggests that \emph{SciQ} is a useful dataset for studying reading
comprehension models in medium-data settings.

\begin{table}[t]
  \centering
  \begin{tabular}{lrr}
    \toprule
    \textbf{Dataset} & \textbf{AS Reader} & \textbf{GA Reader} \\
    \midrule
    4\textsuperscript{th} grade & 40.7\% & 37.6\% \\
    4\textsuperscript{th} grade + SciQ & 45.0\% & 45.4\% \\
    Difference & +4.3\% & +7.8\% \\
    \midrule
    8\textsuperscript{th} grade & 41.2\% & 41.0\% \\
    8\textsuperscript{th} grade + SciQ & 43.0\% & 44.3\% \\
    Difference & +1.8\% & +3.3\% \\
    \bottomrule
  \end{tabular}
  \caption{Model accuracies on real science questions validation set
  when trained on 4\textsuperscript{th} / 8\textsuperscript{th} grade exam questions alone, 
  and when adding \emph{SciQ}.}
  \label{tab:sciq_omnibus_performances}
\end{table}

\textbf{Direct Answer Setting.}  We additionally present a baseline on the direct answer
version of \emph{SciQ}.  We use the Bidirectional Attention
Flow model (BiDAF; ~\newcite{Seo2016Bidirectional}), which recently achieved state-of-the-art results on SQuAD~\cite{rajpurkar2016SQUAD}.  We trained BiDAF on the training portion of
\emph{SciQ} and evaluated on the test set.  BiDAF achieves a 66.7\% exact match and 75.7 F1 score,
which is 1.3\% and 1.6\% below the model's performance on SQuAD.

\subsection{Using \emph{SciQ} to answer exam questions}

Our last experiment with \emph{SciQ} shows its usefulness as training data for models that 
answer real science questions.  We collected a corpus of 4\textsuperscript{th} and
8\textsuperscript{th} grade science exam questions and used the AS Reader and GA Reader to answer
these questions.\footnote{There are approx. 3200 8\textsuperscript{th} grade training
questions and 1200 4\textsuperscript{th} grade training questions.  Some of the questions come from
\url{www.allenai.org/data}, some are proprietary.} Table~\ref{tab:sciq_omnibus_performances} shows model
performances when only using real science questions as training data, and when
augmenting the training data with \emph{SciQ}.  By adding \emph{SciQ}, performance
for both the AS Reader and the GA Reader improves on both grade levels,
in a few cases substantially.  This contrasts with
our earlier attempts using purely synthetic data, 
where we saw models overfit the synthetic data and an overall performance decrease.  Our successful transfer of
information from \emph{SciQ} to real science exam questions shows that the question distribution is similar to that of real science questions.

\section{Conclusion}

We have presented a method for crowdsourcing the creation of multiple choice QA data, with a
particular focus on science questions. 
Using this methodology, we have constructed
a dataset of 13.7K science questions, called \emph{SciQ}, which we release for future research. We
have shown through baseline evaluations that this dataset is a useful research resource, both to
investigate neural model performance in medium-sized data settings, and to augment training
data for answering real science exam questions.

There are multiple strands for possible future work. One direction is a systematic exploration of
multitask settings to best exploit this new dataset.  Possible extensions for the direction of
generating answer distractors could lie in the adaptation of this idea in negative sampling,
e.g.~in KB population.
Another direction is to
further bootstrap the data we obtained to improve automatic document selection, question generation
and distractor prediction to generate questions fully automatically.

\bibliography{emnlp2017}
\bibliographystyle{emnlp_natbib}

\clearpage
\appendix

\newpage
\section{List of Study Books}
\label{sup:books}
The following is a list of the books we used as data source:

\begin{itemize}
\item OpenStax, Anatomy \& Physiology. OpenStax. 25 April 2013\footnote{Download for free at \url{http://cnx.org/content/col11496/latest/}}
\item OpenStax, Biology. OpenStax. May 20, 2013\footnote{Download for free at \url{http://cnx.org/content/col11448/latest/}}
\item OpenStax, Chemistry. OpenStax. 11 March 2015\footnote{Download for free at \url{http://cnx.org/content/col11760/latest/}}
\item OpenStax, College Physics. OpenStax. 21 June 2012\footnote{Download for free at \url{http://cnx.org/content/col11406/latest}}
\item OpenStax, Concepts of Biology. OpenStax. 25 April 2013\footnote{Download for free at \url{http://cnx.org/content/col11487/latest}}
\item Biofundamentals 2.0 -- by Michael Klymkowsky, University of Colorado \& Melanie Cooper, Michigan State                              University\footnote{\url{https://open.umn.edu/opentextbooks/BookDetail.aspx?bookId=350}}
\item Earth Systems, An Earth Science Course on \texttt{\small www.curriki.org}\footnote{\url{http://www.curriki.org/xwiki/bin/view/Group_CLRN- OpenSourceEarthScienceCourse/}}
\item General Chemistry, Principles, Patterns, and Applications by Bruce Averill, Strategic Energy Security Solutions and Patricia Eldredge, R.H. Hand, LLC; Saylor Foundation\footnote{\url{https://www.saylor.org/site/textbooks/General\%20Chemistry\%20Principles,\%20Patterns,\%20and\%20Applications.pdf}}
\item General Biology;  Paul Doerder, Cleveland State University \& Ralph Gibson, Cleveland State University \footnote{\url{https://upload.wikimedia.org/wikipedia/commons/4/40/GeneralBiology.pdf}}
\item Introductory Chemistry by David W. Ball, Cleveland State University. Saylor Foundation \footnote{\url{https://www.saylor.org/site/textbooks/Introductory\%20Chemistry.pdf}}
\item The Basics of General, Organic, and Biological Chemistry by David Ball, Cleveland State University
 \& John Hill, University of Wisconsin
\& Rhonda Scott, Southern Adventist University. Saylor Foundation\footnote{\url{http://web.archive.org/web/20131024125808/http://www.saylor.org/site/textbooks/The\%20Basics\%20of\%20General,\%20Organic\%20and\%20Biological\%20Chemistry.pdf}}
\item Barron's New York State Grade 4 Elementary-Level Science Test, by Joyce Thornton Barry and Kathleen Cahill \footnote{We do not include documents from this resource in the dataset.}
\item Campbell Biology: Concepts \& Connections by Jane B. Reece, Martha R. Taylor, Eric J. Simon, Jean L. Dickey\footnote{We do not include documents from this resource in the dataset.}
\item CK-12 Peoples Physics Book Basic \footnote{\url{http://www.ck12.org/book/Peoples-Physics-Book-Basic/}}
\item CK-12 Biology Advanced Concepts \footnote{\url{http://www.ck12.org/book/CK-12-Biology-Advanced-Concepts/}}
\item CK-12 Biology Concepts \footnote{\url{http://www.ck12.org/book/CK-12-Biology-Concepts/}}
\item CK-12 Biology \footnote{\url{http://www.ck12.org/book/CK-12-Biology/}}
\item CK-12 Chemistry - Basic \footnote{\url{http://www.ck12.org/book/CK-12-Chemistry-Basic/}}
\item CK-12 Chemistry Concepts -- Intermediate  \footnote{\url{http://www.ck12.org/book/CK-12-Chemistry-Concepts-Intermediate/}}
\item CK-12 Earth Science Concepts For Middle School\footnote{\url{http://www.ck12.org/book/CK-12-Earth-Science-Concepts-For-Middle-School/}}
\item CK-12 Earth Science Concepts For High School\footnote{\url{http://www.ck12.org/book/CK-12-Earth-Science-Concepts-For-High-School/}}
\item CK-12 Earth Science For Middle School \footnote{\url{http://www.ck12.org/book/CK-12-Earth-Science-For-Middle-School/}}
\item CK-12 Life Science Concepts For Middle School \footnote{\url{http://www.ck12.org/book/CK-12-Life-Science-Concepts-For-Middle-School/}}
\item CK-12 Life Science For Middle School \footnote{\url{http://www.ck12.org/book/CK-12-Life-Science-For-Middle-School/}}
\item CK-12 Physical Science Concepts For Middle School\footnote{\url{http://www.ck12.org/book/CK-12-Physical-Science-Concepts-For-Middle-School/}}
\item CK-12 Physical Science For Middle School \footnote{\url{http://www.ck12.org/book/CK-12-Physical-Science-For-Middle-School/}}
\item CK-12 Physics Concepts - Intermediate \footnote{\url{http://www.ck12.org/book/CK-12-Physics-Concepts-Intermediate/}}
\item CK-12 People's Physics Concepts \footnote{\url{http://www.ck12.org/book/Peoples-Physics-Concepts/}}
\end{itemize}
CK-12 books were obtained under the Creative Commons Attribution-Non-Commercial 3.0 Unported (CC BY-NC 3.0) License \footnote{\url{http://creativecommons.org/licenses/by-nc/3.0/}}.

\section{Training and Implementation Details}
\textbf{Multiple Choice Reading Comprehension.} During training of the AS Reader and GA Reader, we monitored model performance 
after each epoch and stopped training when the error on the validation set had 
increased (early stopping, with a patience of one). We set a hard limit of ten epochs, but most models reached their peak 
validation accuracy after the first or second epoch. Test set evaluation, 
when applicable, used model parameters at the epoch of their peak validation 
accuracy. We implemented the models in Keras, and ran them with the Theano backend on a Tesla K80 GPU. 

The hyperparameters for each of the models were adopted from previous work. 
For the AS Reader, we use an embedding dimension of 256 and GRU hidden layer 
dimension of 384 (obtained through correspondence with the authors of~\newcite{Onishi2016Who}) and 
use the hyperparameters reported in the original paper \cite{Kadlec2016ASRNetwork} for 
the rest. For the GA Reader, we use three gated-attention layers with the multiplicative 
gating mechanism. We do not use the character-level embedding features or the question-evidence 
common word features, but we do follow their work by using pretrained 100-dimension 
GloVe vectors to initialize a fixed word embedding layer. Between each gated attention 
layer, we apply dropout with a rate of 0.3. The other hyperparameters are the same as 
their original work \cite{Dgingra2016Gated}.

\textbf{Direct Answer Reading Comprehension.} We implemented the Bidirectional Attention Flow model exactly as described in~\newcite{Seo2016Bidirectional} 
and adopted the hyperparameters used in the paper.
\end{document}